\documentclass[preprint,12pt]{elsarticle}
\usepackage{graphicx}
\usepackage{amsmath}
\usepackage{graphics}%
\usepackage{epsfig}
\usepackage{amsfonts}%
\usepackage{amssymb}
\usepackage{amsthm}
\usepackage{lineno}
\newtheorem{theorem}{Theorem}

\journal{Neurocomputing}
\begin{document}

\begin{frontmatter}

\title{Learning Dictionary From Signals under Global Sparsity Constraint}
\author[rvt]{Deyu Meng}
\cortext[cor1]{Corresponding author. Tel: 86 130 3290 4180; Fax: 86
29 8266 8559.} \ead{dymeng@mail.xjtu.edu.cn.}
\author[rvt]{Qian Zhao}
\author[rvt2]{Yee Leung}
\author[rvt]{Zongben Xu}
\address[rvt]{Institute for Information and System Sciences and Ministry of Education Key Lab for Intelligent
Networks and Network Security, Xi'an Jiaotong
University, Xi'an 710049, P.R. China}
\address[rvt2]{Department of Geography \& Resource Management, The Chinese University of Hong
Kong, Hong Kong, P.R. China.}

\begin{abstract}
A new method is proposed in this paper to learn overcomplete dictionary
from signals. Differing from the current methods that enforce uniform
sparsity constraint on the coefficients of each input signal, the proposed method
attempts to impose global sparsity constraint on the coefficient matrix of
the entire signal set.
This enables the proposed method to fittingly assign the atoms of the
dictionary to represent various signals and optimally adapt to the
complicated structures underlying the entire signal set. By virtue of the
sparse coding and sparse PCA techniques, a simple algorithm is designed
for the implementation of the method. The efficiency and the convergence
of the proposed algorithm are also theoretically analyzed. Based on the
experimental results implemented on a series of signal and image data
sets, the capability of the proposed method is substantiated
in original dictionary recovering, signal
reconstructing and salient signal structure revealing.
\end{abstract}


\end{frontmatter}

\section{Introduction}

In recent years, there has been a significant interest in using sparse
representation over a redundant dictionary as a driving force for various
signal processing tasks. All of these applications
capitalize on the fact that salient features in signals can
always be captured by their sparse representations over an appropriate
dictionary. As such, the pre-specified dictionary is crucial to the
success of the sparse representation model in practical applications. Most
conventional studies use the ``off-the-shelf" dictionaries, such as the
wavelet \cite{wav} and DCT bases \cite{DCT}, to build a sparsifying
dictionary based on a mathematical model of the data. Current studies,
however, have demonstrated the advantages of learning an often
overcomplete dictionary matched to signals of interest
\cite{K-SVD}-\cite{Bay}.

The dictionary learning task is mathematically described as follows: For a collection of signals $\textbf{X}=[\textbf{x}_1,\textbf{x}_2,\cdots,\textbf{x}_n]\in
R^{d\times n}$, it is expected to find the dictionary
$\textbf{D}=[\textbf{d}_1,\textbf{d}_2,\cdots,\textbf{d}_m]\in R^{d\times m}$,
composed by a collection of atoms $\textbf{d}_i$ (the atom number $m$ is set
larger than $d$, implying that the dictionary is redundant), through the following
optimization model \cite{efficient}\cite{non-convex}:
\begin{equation}
\underset{\textbf{D},\textbf{A}}{\min}\ \ \frac{1}{n}{\sum}_{i=1}^n\
\left(\frac{1}{2}\|\textbf{x}_i-\textbf{D}\textbf{a}_i\|_2^2 +\lambda
\mathcal{P}(\textbf{a}_i)\right) \tag{$P_{\lambda}$},
\end{equation}
where the vector $\textbf{a}_i$ contains the representation coefficients of
$\textbf{x}_i$. We denote the coefficient matrix as
$\textbf{A}=[\textbf{a}_1,\textbf{a}_2,\cdots,\textbf{a}_n]\in R^{m\times n}$. The objective
function of ($P_{\lambda}$) involves two elements in the dictionary
learning task: the expression error term,
$\frac{1}{2}\|\textbf{x}_i-\textbf{D}\textbf{a}_i\|_2^2$, and the sparsity controlling term,
$\mathcal{P}(\textbf{a}_i)$, with respect to each input signal $\textbf{x}_i$. The
most widely utilized functions of $\mathcal{P}(\textbf{a}_i)$ include the
$l_0$ penalty $\|\textbf{a}_i\|_0$ and the $l_1$ penalty $\|\textbf{a}_i\|_1$. The other two generally utilized models for the dictionary task are \cite{K-SVD}\cite{K-SVD2}:
\begin{equation}
\underset{\textbf{D,A}}{\min}\left\Vert \textbf{X} - \textbf{DA}\right\Vert _{F}^2\text{ \ \ s.t. \ \
}\mathcal{P}(\textbf{a}_i)\leq k,\ \ \forall \ \ 1\leq i\leq n, \tag{$P_{k}$}
\end{equation}
and
\begin{equation}
\underset{\textbf{D,A}}{\min}\ {\sum}_{i=1}^n\
\mathcal{P}(\textbf{a}_i) \text{ \ \ s.t. \ \ }\left\Vert \textbf{x}_i - \textbf{D}\textbf{a}_{i}
\right\Vert _{2}\leq \epsilon, \ \ \forall \ \ 1\leq i\leq n,
\tag{$P_{\epsilon}$}
\end{equation}
where the notion $\|\cdot\|_F$ stands for the Frobenius norm. The tunable
parameters $\lambda$, $k$ and $\varepsilon$ in the models
($P_{\lambda}$), ($P_k$) and ($P_{\epsilon}$) play an important role
in the model performance. They intrinsically control the compromise
between the expression error and the sparsity
of the representation coefficients.

It should be noted that a uniform parameter $\lambda$, $k$ or
$\varepsilon$ formulated for the entire signal set is specified in the
current model ($P_{\lambda}$), ($P_k$) or ($P_{\epsilon}$), respectively.
Such formulation facilitates the parameter selection and algorithm
construction of the model. The signals in applications, however, are
always of varying interior structures. On one hand, some signals may be
composed of complicated features and need to be very densely represented
under the dictionary; while some might be of very simple structure and can
be precisely represented with very sparse coefficient vectors. On the
other hand, some signals may seriously deviate from the original due to
the analog-to-digital conversion errors or transmission through noisy
channels, while some may be totally clean samples. This means that it is better to vary the parameter $\lambda$, $k$ or $\varepsilon$ with respect to
different signals to make the dictionary learning model adaptive to the
intrinsic structures underlying the entire signal set. The uniform
specification of the sparsity penalty $\lambda$, the maximal
representation sparsity $k$, or the minimal representational error
$\varepsilon$ of the conventional model ($P_{\lambda}$), ($P_k$) or
($P_{\epsilon}$), respectively, thus possibly
conducts unstable performance of the model in applications.

The purpose of this paper is to formulate a new dictionary learning model,
which is with simple form while will not impose uniform penalty or constraint on each signal like the
conventional methods. Instead, the model will specify a global sparsity
constraint on the coefficients of the entire signal set, so that it will adaptively tune the
representation sparsity of diverse signals and properly reveal the
intrinsic structures underlying the entire signal set. An efficient
algorithm is specifically designed for the proposed model. It is
efficient, convergent and easy to be implemented. By a series of experiments, it is verified that
the proposed algorithm, in comparison with the current dictionary learning
methods, can not only deliver more faithful dictionary underlying the
input signals, but also can more precisely recover the original signals.
Besides, the intrinsic structure underneath the
entire signal set can be impressively depicted via the different
representation sparsities of the signals under the learned dictionary.

In what follows, related work in the literature is first reviewed in
Section II. Details of our algorithm and its basic model are then
presented in Section III. The
experimental results are given in Section IV for substantiation and
verification. The paper is then concluded with a summary and outlook for
future research. Throughout the paper, we denote matrices, vectors and scalars by the
upper-case bold-faced letters, lower case bold-faced letters and
lower-case letters, respectively.

\section{Related work}

Using sparse representations of signals under an appropriately specified
redundant dictionary has advanced multiple signal processing tasks, and
has drawn much attention recently. In the conventional
studies, the ``off-the-shelf" dictionaries have always been employed in
various applications. The typical ones include the Fourier \cite{Fourier},
the wavelet \cite{wav1}-\cite{wav5} and the DCT bases \cite{DCT}.
These bases have been successfully applied to many practical problems.

Recent research, however, has demonstrated the significance of learning an
overcomplete dictionary, instead of a fixed one, matched to the signals of
interest. Various algorithms along this line have been formulated in
recent years. For example, the algorithm proposed by Olshausen and Field
\cite{nature} can find sparse linear codes for natural scenes. The
dictionary composed by these codes complies with the intrinsic features of
the localized, oriented, bandpass receptive fields of the neurons of the
primary visual cortex. The method of optimal directions (MOD), proposed by
Engan et al. \cite{MOD}, is also an appealing dictionary training
algorithm. It improves the efficiency of the work by Olshausen and Field
\cite{nature} both in the sparse coding and dictionary updating stages.
Through generalizing the K-means clustering process to alternate between
sparse coding and dictionary updating, Aharon et al. \cite{K-SVD,K-SVD2}
designed the K-SVD method. There are two versions of the method: one achieves
sparse signal representations under strict sparsity constraint
(corresponding to the model ($P_k$), and is thus denoted as K-SVD$_{P_k}$)
\cite{K-SVD}, and the other calculates the dictionary by allowing a bounded
representation error for each signal (corresponding to the model
($P_\varepsilon$), and is thus denoted as K-SVD$_{P_\varepsilon}$)
\cite{K-SVD2}. The method has empirically shown the state-of-the-art
performance in some image processing applications. Some methods have
further been constructed to improve the efficiency of the dictionary
learning problem: Mairal et al. \cite{online} proposed an online model to
efficiently solve the dictionary learning problem; Jenatton et al.
\cite{structure} used a tree structured sparse representation to give a
linear-time computation of the problem; and Lee et al. \cite{efficient}
designed an algorithm to speedup the sparse coding stage of the problem,
allowing it to learn larger sparse codes than other algorithms. Recently,
some algorithms have also been developed to extend the capability of
dictionary learning based on some specific motivations. For instance, Shi
et al. \cite{non-convex} developed an algorithm for dictionary learning
with non-convex while continuous minimax concave penalty; Mairal et al. \cite{super}
established a discriminative approach, instead of the purely
reconstructive methods, to build a dictionary. All of the aforementioned
methods are addressed to the models ($P_{\lambda}$), ($P_k$) and
($P_{\epsilon}$) introduced in Section I.

For many real signal processing applications, however, these models cannot
fully tally with the practical signals possessing intrinsic complicated
structures, such as those with different content of features, or with
spatial and/or spectral non-uniform noises. Along this line of research,
Mairal et al. \cite{K-SVD3} addressed the case of removing nonhomogeneous
white Gaussian noise of images, while apriori knowledge of the noise
deviation at each pixel of the objective images has to be pre-assumed.
Such elaborate information about noise, however, generally cannot be
attained in practice. Spielman et al. \cite{colt} proposed a method called ERSuD, which is also based on the global $l_0$
sparsity of a matrix. However, this method is applicable only when the dictionary matrix is square and invertible, which
generally does not hold in real dictionary learning applications. Very recently, Zhou et al. \cite{Bay2,Bay} designed
a nonparametric Bayesian method, called the beta process factor analysis
(BPFA) method, for dictionary learning. The method can learn a sparse
dictionary in situ for signals with spatially non-uniform noises, without
having to know the apriori noise information. The method is thus used as
one of the methods for comparison in our experiments.

In this paper, we propose a new dictionary learning method to enhance the
capability of dictionary learning by imposing a global
sparsity constraint on the coefficients of all training signals to enable adaptive atom
assignment to individual signals based on their intrinsic structures.
It should be noted that the concept of global-sparsity-constraint has also been involved in some other problems, such
as image decomposition \cite{G1} and magnetic resonance imaging \cite{G2}, where this idea was verified to be beneficial to achieve a global consistency of data structures. In machine learning area,
there are also some literatures modeling the matrix with $l_1$-norm global sparsity \cite{G3}-\cite{G5}.
Here we first introduce the $l_0$-norm global sparsity in dictionary learning problem.

\section{Dictionary learning under global sparsity constraint: model and algorithm}

\subsection{Model: From local to global constraint on sparse representation}

\begin{figure}[t]
\begin{center}
\scalebox{0.45}[0.45]{\includegraphics[bb=195 330 390 550]{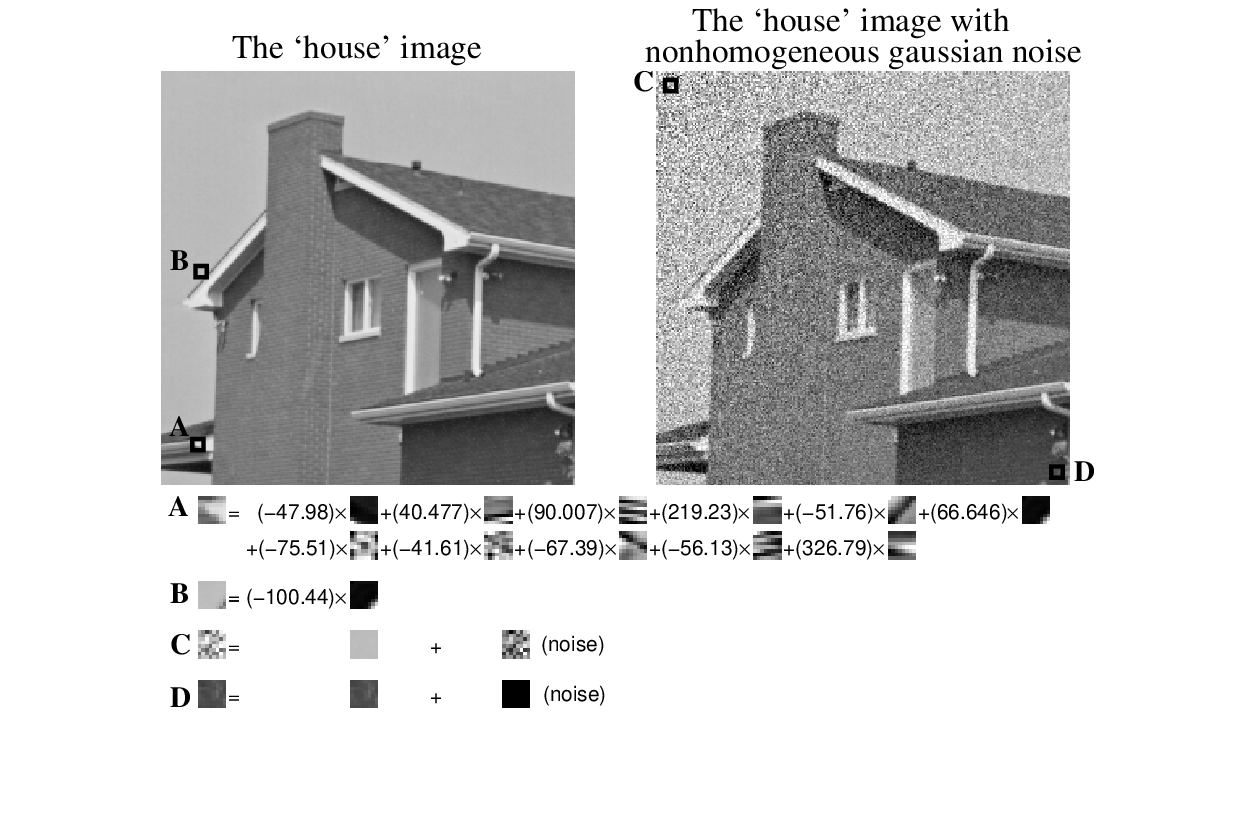}}
\end{center}
\caption{The left figure is the ``house" image, and the right figure is the
same image mixed with nonhomogeneous Gaussian noise. A,B,C,D are four
patches cut from the two images. The upper two expressions
at the bottom of the figure show the atoms and the coefficients utilized
to sparsely represent the patches A and B over the dictionary learned
from the algorithm proposed in Section 3.2, respectively. The lower two
ones demonstrate the groundtruth of the noise separated from the patches
C and D, respectively.} \label{f1}
\end{figure}

The current dictionary learning models, namely ($P_{\lambda}$), ($P_k$)
and ($P_{\epsilon}$), enforce uniform sparsity control parameter,
including sparsity penalty $\lambda$, sparsity constraint $k$ or
representation error bound $\epsilon$, for each involved signal. However,
there are often counterexamples to such formulation in real applications,
especially for signals embedded with intrinsic heterogeneous sparsity
structures. We take the image case as an instance, in which the input
signal set corresponds to the small local patches of the image in
consideration. On one hand, the local parts of a real image may contain
very different capacities of meaningful features, e.g., the region full of
patterns with abundant textures, as compared to the area located at the
background with small grayscale variations. In such case, smaller sparsity
penalty $\lambda$ of the optimization model ($P_{\lambda}$) or larger representation sparsity
$k$ of the optimization model ($P_k$), should be preset for the local patches located at the
former region so that more atoms of the dictionary can be assigned to
them. This phenomenon can be easily understood via the representations of
the patches A and B located at the house eave and the background parts of
the ``house" image, respectively, as shown in Figure \ref{f1}. On the
other hand, the real noise mixed in the image is often of significant
statistical heteroscedasticity. This means that the extents of noise
corruption in various parts of the image, such as the patches C and D in
Figure \ref{f1}, may be significantly different. It is easy to see that
the patch C is highly corrupted by noise while D is almost clean and
contains essentially no noise. Larger representation error bound
$\epsilon$ of the optimization model ($P_{\epsilon}$) should then be set for the former patches
so that the model can properly capture the variation of noise corruption
across the image. It is thus foreseeable that the performance of the
current dictionary learning models can be substantially enhanced by
adaptively regulating the sparsity control parameter(s) with respect to
the underlying structural characteristics of the entire signals.

Based on the above rationale, we reformulate the model for dictionary
learning into the following global-sparsity-constraint form:
\begin{equation}
\underset{\textbf{D,A} }{\text{min} }\left\Vert\textbf{X}-\textbf{DA}\right\Vert _{F}^2\text{ \ \
s.t. \ \ } \left\Vert \textbf{A}\right\Vert _{0}\leq K, \tag{$P_{K}$}
\end{equation}
where $K$ is the maximal size of the non-zero entries of the coefficient
matrix $\textbf{A}$ (i.e., the combination of the $l_0$ sparsities of all signals), and $\|\textbf{A}\|_0$
counts the nonzero entries of the matrix $\textbf{A}$. Differing from the current models in which uniform sparsity
constraint is imposed on the coefficient vector for each input signal, our proposed model capitalizes
on the global sparsity constraint superimposed upon the coefficient matrix for the entire signal set.
This formulation enables the model to adaptively assign different number
of atoms, $k_i$, of the dictionary to represent each signal $\textbf{x}_i$
according to its intrinsic structure. This can be easily understood
through the following equivalent reformulation of ($P_{K}$):
\begin{equation}
\underset{\textbf{D},\textbf{A}, \{k_i\}_{i=1}^n}{\text{min} }\left\Vert \textbf{X}-\textbf{DA}\right\Vert
_{F}^2\text{ \ \ s.t. \ \ } \left\Vert \textbf{a}_i \right\Vert _{0}\leq k_i,\
1 \leq i \leq n, \ \ {\sum}_{i=1}^n k_i = K
.\label{loc}
\end{equation}
In specific, for signals containing different capacities of features, more
non-zero atoms will be assigned to represent more complex signals by the
proposed model; and for signals corrupted by the heterogeneous noises, the
distribution of the non-zero entries of the coefficient matrix tends to be
optimally balanced among the entire signal set, and the sparse
representations of the signals over the dictionary attained by the
proposed model will possibly reveal the major information (the original
signals) while eliminate the minor (the mixed noises) at the global scale.
The dictionary learning model ($P_{K}$) is thus expected to outperform the
current models.

We now construct an efficient algorithm for solving ($P_{K}$).

\subsection{Algorithm: Iteratively updating columns and rows of coefficient matrix}

\begin{figure*}
\begin{center}
\scalebox{0.35}[0.35
]{\includegraphics[bb=500 150 380 390]{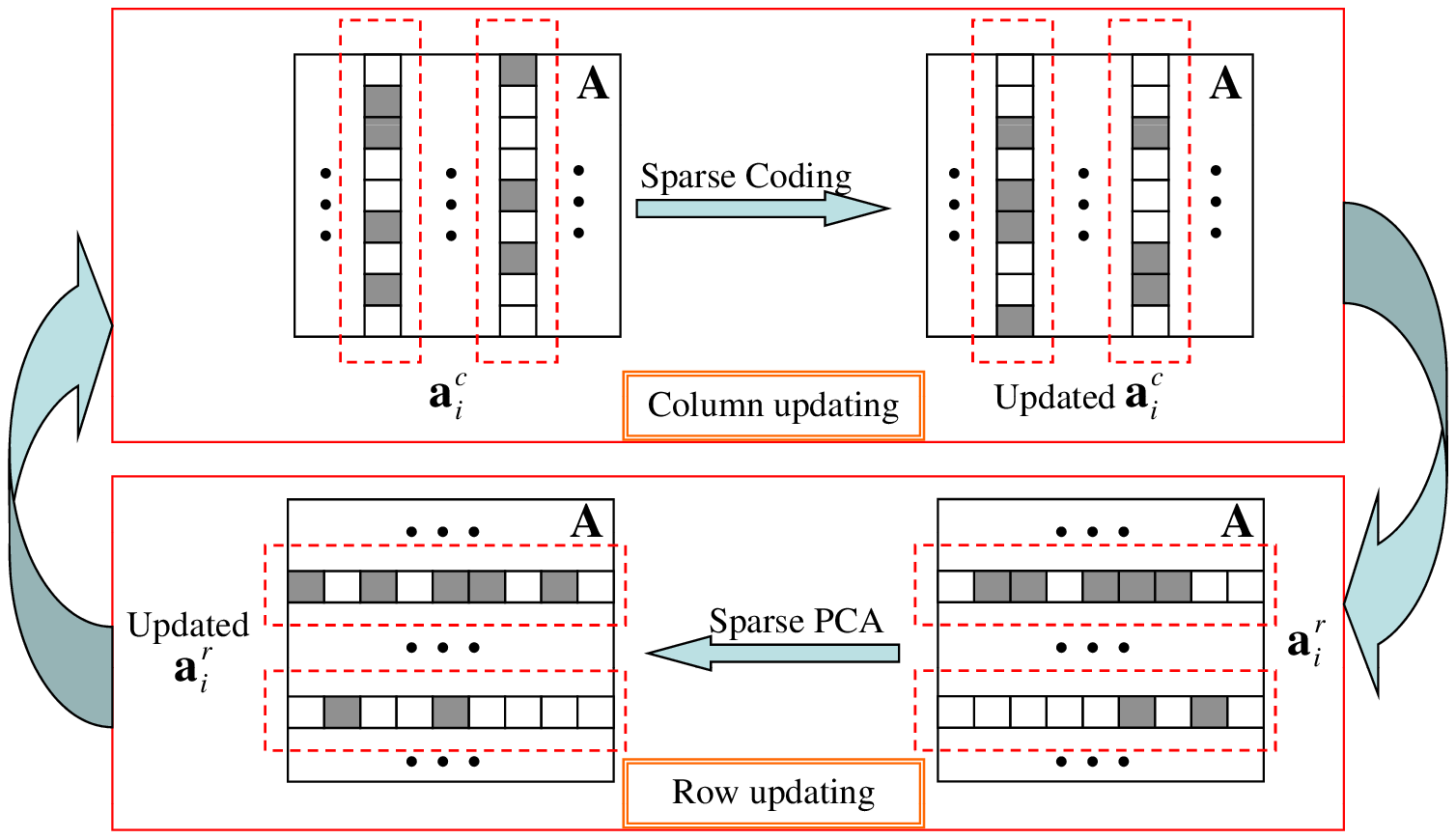}}
\end{center}
\caption{Graphical presentation of the iteration process between the column updating step and
the row updating step in the proposed algorithm. The upper panel
shows that in the column updating step, each column vector
$\textbf{a}^c_{i}$ of $\textbf{A}$ is updated, with the number of its non-zero
entries fixed while the column positions of these non-zero
elements are optimally relocated in an adaptive way. The lower panel demonstrates that
in the row updating step, each row vector $\textbf{a}^r_{i}$
of $\textbf{A}$ is updated, with its sparsity
fixed while its non-zero elements are optimally adapted to the
proper row positions.} \label{f}
\end{figure*}

The main idea of our algorithm is to iteratively update the column and row
vectors of the coefficient matrix $\textbf{A}$ to approach the solution to
($P_{K}$). Denote the column and row vectors of the coefficient matrix $\textbf{A}$ ($\in R^{m\times n}$)
as $[\textbf{a}^c_{1},\textbf{a}^c_{2},\cdots,\textbf{a}^c_{n}]$ and
$[\textbf{a}^r_{1},\textbf{a}^r_{2},\cdots,\textbf{a}^r_{m}]$,
respectively. The column updating step is to update each column vector
$\textbf{a}^c_{i}$ ($i=1,2,\cdots,n$) of $\textbf{A}$, with the number of its non-zero
entries, $k_i^c$, fixed, while optimally relocate the column positions of these non-zero
elements in an adaptive way (i.e., $k_i^r$s will be varied after this step), as graphically depicted
in the upper of Figure \ref{f}.
By decoupling the
model ($P_{K}$), the corresponding task is to solve the following
optimization model for each $\textbf{a}^c_{i} $ ($i=1,\cdots,n$):
\begin{equation}
\underset{\textbf{a}^c_{i} }{\text{min} }\left\Vert \textbf{x}_i-\textbf{D}\textbf{a}^c_{i}
\right\Vert _{2}^2\text{ \ \ s.t. \ \ } \left\Vert\textbf{a}^c_{i}\right\Vert
_{0}\leq k_i^c. \label{e1}
\end{equation}
The row updating step is to update each row vector $\textbf{a}^r_{i}$
($i=1,2,\cdots,m$) of $\textbf{A}$, with the sparsity $k_i^r$ of $\textbf{a}^r_{i}$
fixed, while optimally adapt its $k_i^r$ non-zero elements to the
proper row positions (i.e., $k_i^c$s will be changed after this step), as shown in the lower of Figure \ref{f}.
Since the atom $\textbf{d}_i$ of the dictionary $\textbf{D}$
one-to-one corresponds to $\textbf{a}^r_{i}$ in the sense that
$\textbf{DA}={\sum }_{i=1}^m%
\textbf{d}_{i}\left( \textbf{a} _{i}^{r}\right) ^{T}$, it is also simultaneously
updated in this step together with $\textbf{a}^r_{i}$. The corresponding
optimization model is of the following form for each $\textbf{a}^r_{i}$
($i=1,\cdots,m$):
\begin{equation}
\underset{\textbf{a}^r_i,\textbf{d}_i }{\text{min}}\left\Vert
\textbf{E}_i-\textbf{d}_i(\textbf{a}^r_i)^T\right\Vert _{F}^2 \text{\ \ s.t. }\ \
\left\Vert\textbf{a}^r_i\right\Vert _{0}\leq k_i^r, \ \ \textbf{d}_i^T\textbf{d}_i=1, \label{e2}
\end{equation}
where $\textbf{E}_i=\textbf{X}-{\sum }_{j\neq i}%
\textbf{d}_{j}\left( \textbf{a} _{j}^{r}\right) ^{T}$ stands for the representation
error of all considered signals with the effect of the $i$-th atom $\textbf{d}_i$
removed. It should be noted that the
sparsity $k_i^c$ of each $\textbf{a}^c_{i}$ and $k_i^r$ of each $\textbf{a}^r_{i}$ are dynamically and adaptively adjusted during the iterations between the column updating and the row updating steps of the proposed algorithm.

The algorithm can then be summarized as Algorithm 1.

\begin{table*}
\small
\begin{tabular}{l}
\\ \hline
\textbf{Algorithm 1} Algorithm for dictionary learning under global
sparsity constraint (GDL)
\\ \hline
Given: The input data $\textbf{X}=[\textbf{x}_1,\cdots,\textbf{x}_n]\in R^{d\times n}$, the global sparsity $K$ \\
Execute:
\\
\ \ \ \ \ 1. Initialize the dictionary $\textbf{D}\in R^{d\times m}$ and the
coefficient matrix $\textbf{A}\in R^{m\times n}$\\
\ \ \ \ \ \ \ \ \ with sparsity $K$, respectively.
\\
\ \ \ \ \ 2. Repeat
\\
\ \ \ \ \ \ \ \ \ 2.1 (Column updating). Update the column vector
$\textbf{a}^c_{i}$ of $\textbf{A}$ by solving (\ref{e1})\\
\ \ \ \ \ \ \ \ \ \ \ \ \ \ for each $i=1,\cdots,n$.
\\
\ \ \ \ \ \ \ \ \ 2.2 (Row updating). Update the row vector $\textbf{a}^r_{i}$
of $\textbf{A}$ and the atom $\textbf{d}_i$ of \\
\ \ \ \ \ \ \ \ \ \ \ \ \ \ $\textbf{D}$ by solving (\ref{e2}) for each
$i=1,\cdots,m$.
\\
\ \ \ \ \ \ Until the termination condition is
satisfied\\
Return:  the solution $\textbf{D,\ A}$ of ($P_K$).
\\\hline
\end{tabular}
\end{table*}

Now the question is how to efficiently solve the optimization models
(\ref{e1}) and (\ref{e2}). For (\ref{e1}), it is actually the well known
$l_0$-norm model of sparse coding and multiple effective algorithms have
been investigated to solve this model. The typical ones
include the thresholding methods, e.g., the hard algorithm \cite{hard},
and the greedy methods, e.g., the OMP algorithm \cite{OMP}.

For (\ref{e2}), we give the following theorem:
\begin{theorem}
The optimum of the optimization model (\ref{e2}) can be attained by solving the
optimization model:
\begin{equation}
\underset{\textbf{w} }{\text{max}}\ \textbf{w}^T\textbf{E}_i^T\textbf{E}_i\textbf{w}\ \ \ \ \  \text{ s.t. }
\textbf{w}^T\textbf{w}=1,\left\Vert \textbf{w}\right\Vert _{0}\leq k_i^r, \label{e3}
\end{equation}
in the sense of
\begin{equation}
\widehat{\textbf{d}}_i= \frac{\textbf{E}_i\widehat{\textbf{w}}}{\left\Vert \textbf{E}_i\widehat{\textbf{w}}\right\Vert
_{2}}, \widehat{\textbf{a}}^r_i= \left\Vert \textbf{E}_i\widehat{\textbf{w}}\right\Vert
_{2}\widehat{\textbf{w}}, \label{e4}
\end{equation}
where $\widehat{\textbf{d}}_i$, $\widehat{\textbf{a}}^r_i$ are the optimum of
(\ref{e2}), and $\widehat{\textbf{w}}$ is the optimum of (\ref{e3}).
\end{theorem}

The proof of Theorem 1 is given in the appendix.

It is very interesting that (\ref{e3}) is just the sparse principal
component analysis (sparse PCA) model \cite{SPCA}, which has been
thoroughly investigated in the last decade \cite{SPCA}-\cite{EMPCA}. Many
efficient algorithms have been constructed for solving the model,
including SPCA \cite{SPCA}, GPower \cite{Gpower}, sPCA-rSVD \cite{rSVD}
etc..

Thus, both models (\ref{e1}) and (\ref{e2}) can be efficiently solved,
i.e., both of the column and row updating steps of the proposed algorithm
can be effectively implemented, by employing the existing methods in
sparse coding and sparse PCA, respectively.

The remaining issues are then how to appropriately specify the initial
dictionary $\textbf{D}$ and the coefficient matrix $\textbf{A}$ in step 1, and when to
terminate the iterative process in step 2 of the proposed algorithm. In
our experiments, $\textbf{D}$ and $\textbf{A}$ were simply initialized with data signals and
sparse matrix with $K$ non-zero elements, whose positions are randomly
generated in the matrix, respectively. By counting the nonzero element numbers of each column vector $\textbf{a}_i^c$ and each row vector $\textbf{a}_i^r$
of the initialized $\textbf{A}$, the sparsity $k_i^c$ and $k_i^r$ are simultaneously specified.
As for the termination of the algorithm, since the entire representation error of signals, i.e., the
objective function of ($P_K$), decreases monotonically under the fixed
global sparsity constraint throughout the iterative process, the algorithm can be reasonably terminated when the
decrease in value of the objective function is smaller than some preset
small threshold, or the process has reached the pre-specified number of
iterations.

As for the convergence of the proposed algorithm,
under the assumption that the models (\ref{e2}) and (\ref{e3}) can be
precisely solved, each of the updating iterations in step 2 monotonically
decreases the objective function of (\ref{e1}), i.e., the total representation error $\left\Vert \textbf{DA} -\textbf{X}\right\Vert
_{F}^2$, under the guarantee that the constraint $\left\Vert \textbf{A}\right\Vert
_{0}\leq K$ is consistently held. Since this objective function is also lower bounded ($\geq 0$),
the algorithm is guaranteed to be convergent.
Although the above claim depends on the success of the sparse coding and sparse PCA
techniques used to approximate the solutions to (\ref{e2}) and
(\ref{e3}), respectively, the algorithms employed on the tasks always performed
well in our experiments and can empirically generate a rational solution
of (\ref{e4}) after multiple iterations, as demonstrated in the next section.

\section{Experimental results}

To test the effectiveness of the proposed algorithm on dictionary
learning, it was applied to a series of synthetic signals and real images
for substantiation. The results are summarized in the following
discussion.  All programs were implemented on the Matlab 7.0 platform.

\subsection{Synthetic signal experiments with homogeneous noise}

We first apply the proposed algorithm to synthetic signal data to test
whether the algorithm can recover the generating dictionary and
reconstruct the original signals.
Two series of experiments were implemented, each involving $11$
sets of signals. Each signal set contained $1500$ $20$-dimensional
signals, denoted as $\textbf{X}=[\textbf{x}_1,\textbf{x}_2,\cdots,\textbf{x}_{1500}]\in R^{20\times 1500}$,
which were created by a linear combination of a pre-specified dictionary
$\textbf{D}=[\textbf{d}_1,\textbf{d}_2,\cdots,\textbf{d}_{50}]\in R^{20\times 50}$ and representation
coefficients $\textbf{A}=[\textbf{a}_1,\textbf{a}_2,\cdots,\textbf{a}_{1500}]\in R^{50\times
1500}$, and mixed with different extents of homogeneous Gaussian white
noise. The entries of each dictionary $\textbf{D}$ were first generated by random
sampling, and each column (atom) $\textbf{d}_i$ of $\textbf{D}$ was then divided by its
$l_2$-norm for normalization. For the first series of experiments, each
column $\textbf{a}_i$ of $\textbf{A}$ contained $3$ non-zero elements with randomly
chosen values and locations. For the second series of experiments, each
coefficient matrix $\textbf{A}$ consisted of $4500$ randomly valued and located
non-zero entries. Added to the $11$ signal sets in both series of
experiments were homogeneous Gaussian noises with standard deviations $\sigma$
varying from $0$ to $0.1$ with interval $0.01$. The corresponding SNR values of these
signal sets ranged from infinity to around $10$\footnote{The signal set mixed with Gaussian noise with
deviation $0$ means that the set is clean and contains no noise. The
corresponding value of SNR is infinite.}. It should be indicated that for
the first experimental series, the signals in each experiment were of
similar representation sparsity over the preset dictionary, which complies
with the preassumption of the model ($P_k$); and for both series of
experiments, the signals were corrupted with homogeneous noises, which
tallies with the preassumption underlying the model ($P_{\varepsilon}$).

Five of the current dictionary learning methods, including the MOD \cite{MOD}, K-SVD$_{P_k}$
\cite{K-SVD}, K-SVD$_{P_\varepsilon}$ \cite{K-SVD2}, Efficient \cite{efficient} and
BPFA \cite{Bay2} methods were applied to these signal sets for comparison. The
dictionaries of the first four methods, as well as the proposed GDL method,
were initialized as the randomly selected signals from the input set, and
the initialization of the BPFA method was based on the singular value
decomposition technique \cite{Bay2}. Since both MOD and
K-SVD$_{P_\varepsilon}$ need the apriori deviation of the noise mixed in
the signals to preset the representation error parameter, we directly used
the groundtruth information to optimally specify the parameter value
\cite{K-SVD2}. For the K-SVD$_{P_k}$ and Efficient methods, the sparsity constraint parameters $k$ and $\lambda$
were specified by running the method $5$ times on each
signal set under different parameter values, and selecting the best one as the final
output. All parameters involved in the BPFA were automatically inferred by
using a full posterior on the model
\cite{Bay2}. For the proposed GDL method, the global sparsity $K$ was set
as $4500$ in all experiments. The results of K-SVD$_{P_k}$, K-SVD$_{P_\varepsilon}$,
MOD, Efficient and GDL were
attained after $100$ iterations, and those of BPFA were obtained after
$1000$ iterations of Gibbs sampling.

Two criteria are utilized to assess the performance of the employed
methods for dictionary learning. The first is computed by sweeping through
each atom of the generating dictionary and checking whether it is
recovered by the dictionary attained by a utilized method via the
following formula \cite{K-SVD}:
\begin{equation}
1-|\textbf{d}_i^T\widehat{\textbf{d}}_i|, \label{E5}
\end{equation}
where $d_i$ is the atom of the original dictionary and $\widehat{\textbf{d}}_i$ is
its corresponding closest atom in the recovered dictionary. If the
value of (\ref{E5}) was less than $0.01$, then it was considered as a
success. The rate of the successfully recovered atoms in the generating
dictionary (called the dictionary recovery rate, or DR in brief) is then
taken as the first criterion, which evaluates the capability of the method
in delivering the original dictionary beneath the input signals. The
second criterion is the mean of the standard deviations of the
reconstructed signals from the original signals (called the representation
error, or RE briefly). This value assesses the performance of the method
in recovering the input signals.

\begin{figure*}
\begin{center}
\scalebox{0.4}[0.35]{\includegraphics[bb=220 310 380 510]{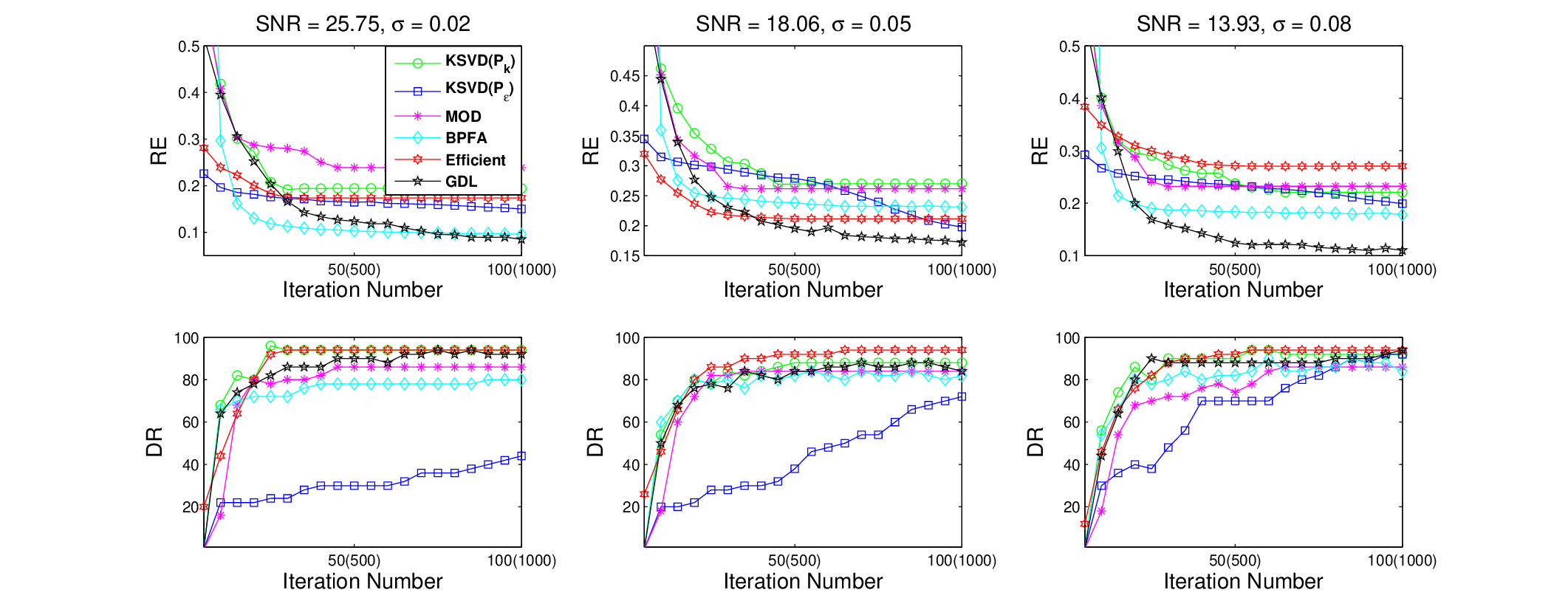}}
\end{center}
\caption{The upper panels: The RE curves of the K-SVD$_{P_k}$,
K-SVD$_{P_{\varepsilon}}$, MOD, BPFA, Efficient and the proposed GDL methods in the
iterative processes of three of the first series of experiments. The lower
panels: the corresponding DR curves. It should be noted that the BPFA method implements
$1000$ iterations of Gibbs sampling, while other methods run $100$ iterations.
} \label{f2}
\end{figure*}

\begin{figure*}
\begin{center}
\scalebox{0.4}[0.35]{\includegraphics[bb=220 300 380 510]{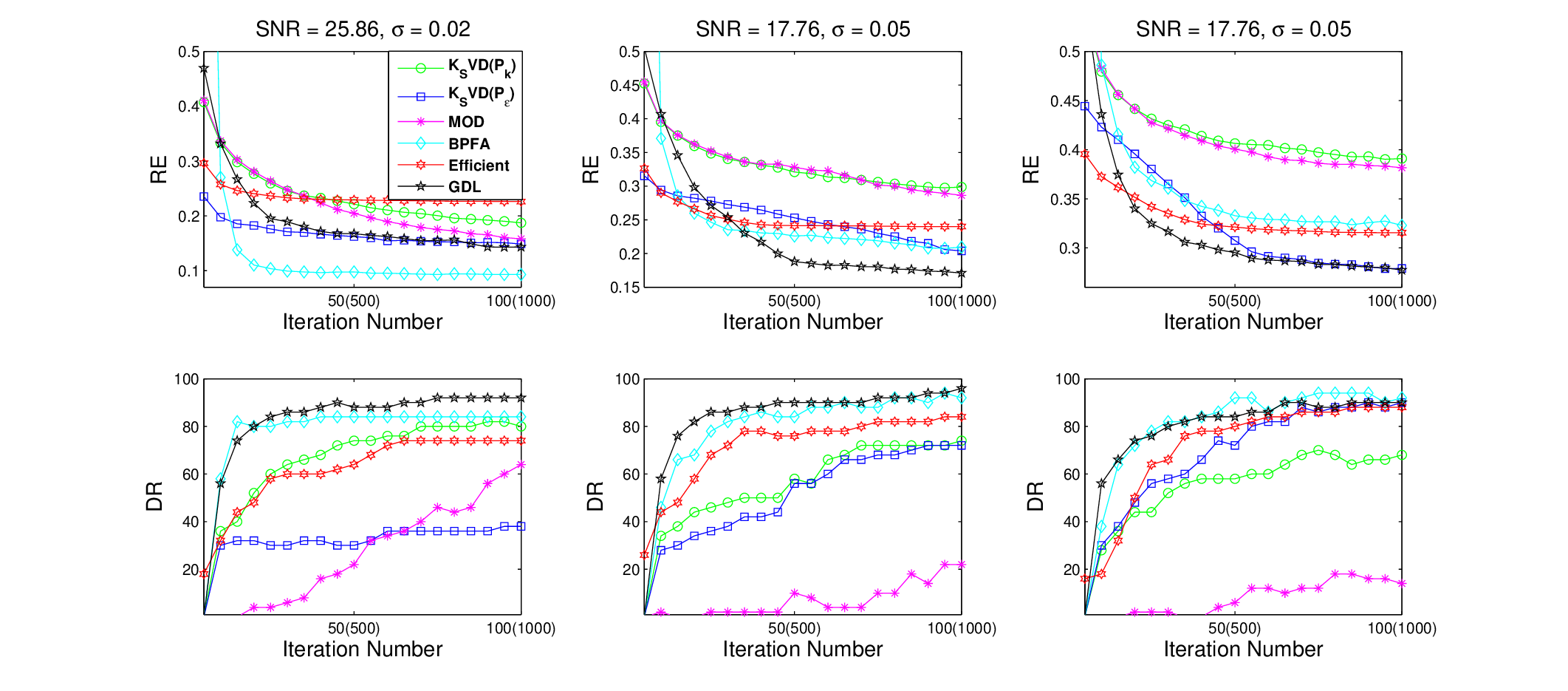}}
\end{center}
\caption{The upper panels: The RE curves of the six utilized methods in the
iterative processes of three of the second series of experiments. The
lower panels: the corresponding DR curves.} \label{f3}
\end{figure*}

\begin{figure*}
\begin{center}
\scalebox{0.4}[0.3]{\includegraphics[bb=220 400 380 450]{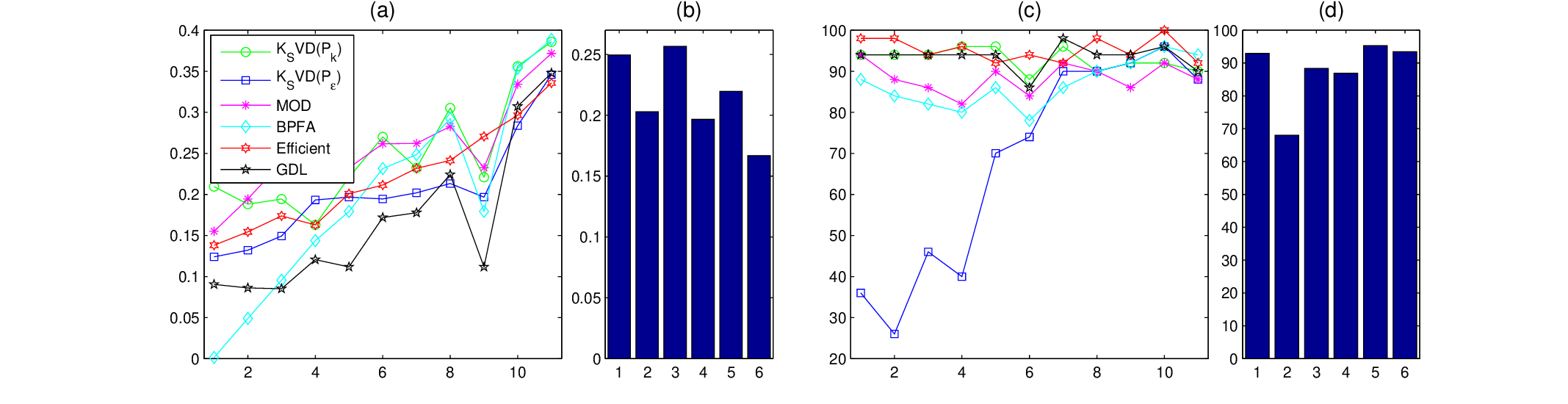}}
\end{center}
\caption{(a)(c): the final RE and DR values obtained by the six utilized methods
in $11$ experiments of the first series. (b)(d): The mean RE and DR values
obtained by the six methods in the first series of experiments. The
numbers $1$-$6$ in the horizonal axis stand for the K-SVD$_{P_k}$,
K-SVD$_{P_{\varepsilon}}$, MOD, BPFA, Efficient and GDL methods, respectively.} \label{f4}
\end{figure*}

\begin{figure*}
\begin{center}
\scalebox{0.4}[0.3]{\includegraphics[bb=220 400 380 540]{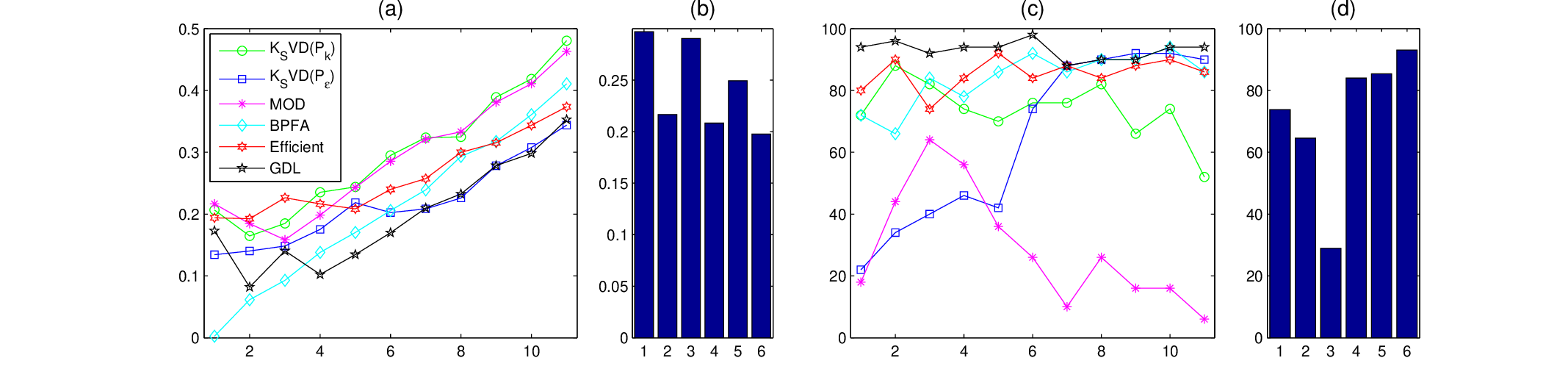}}
\end{center}
\caption{(a)(c): the final RE and DR values obtained by the six utilized methods
in $11$ experiments of the second series. (b)(d): The mean RE and DR
values obtained by the six methods in the second experimental series.} \label{f5}
\end{figure*}

In all of the implemented experiments, the DR and RE values in the
iterative processes of the five current methods and the proposed GDL method
were recorded. The upper and lower panels of Figure \ref{f2} depict the RE
and DR curves of the six methods in the iterative processes of three of
the first series of experiments, respectively. Figure \ref{f3} shows the
corresponding results in three cases of the second series of experiments.
Panels (a) and (c) of Figure \ref{f4} show the final RE and DR values of the
six methods in $11$ experiments of the first series, respectively. For
easy comparison, panels (b) and (d) of the figure display the mean values
of RE and DR of the six methods as vertical bars. Figure \ref{f5} depicts
the cases of the second series of experiments.

It can be easily observed from the upper panels of Figures \ref{f2} and
\ref{f3} that the RE values obtained by the proposed method tend to
decrease monotonically throughout the iterative process. Besides, after
around $20$ iterations, the GDL method attains the smallest or the second smallest RE values among the six methods.
In the final output, the GDL method also achieves the comparatively small RE values in all of the experiments,
as shown in Figure \ref{f4}(a) and Figure \ref{f5}(a). On the average,
the proposed algorithm outperforms the other five methods in both series
of experiments, as shown in \ref{f4}(b) and Figure \ref{f5}(b). This
demonstrates the excellent capability of the proposed algorithm in
reconstructing the input signals.

Furthermore, from the lower panels of Figures \ref{f2} and \ref{f3}, it can
be observed that the DR curves of the proposed method tend to increase
monotonically in all experiments, and the method obtains the largest or the second largest values among the six
methods after $60$ iterations. Moreover, by observing Figure \ref{f4}(c) and
Figure \ref{f5}(c), the proposed algorithm apparently yields the most stable
DR values among the six methods, especially the second series of experiments, where the
input signals are with heterogeneous sparsity structures. The proposed method
successfully detects more than 85\%
atoms of the original dictionary in each of the experiments, and achieves the second largest average DR value
in the first experimental series (only unsubstantially smaller than the Efficient method), and the largest in the second, as shown in Figure \ref{f4}(d) and Figure
\ref{f5}(d), respectively. This substantiates the good capability of the GDL algorithm
in recovering the original dictionary.


In the next section, we further verify the effectiveness of the proposed
method on image reconstruction from data with more complex intrinsic
structures and more complicated nonhomogeneous noises.

\subsection{Real image experiments with nonhomogeneous noise}

A series of test images of $512\times 512$ or $256\times 256$ pixels were
utilized for the image reconstruction problems. These images were
generated by combining $6$ gray-scale images, all of which are widely used
in the image processing literature \cite{6image}, with different levels of
nonhomogeneous noise. In our experiments,
four types of nonhomogeneous noise were employed for each image, constituting four
series of experiments listed as follows.

\emph{Experiment 1 (E1): Nonhomogeneous Gaussian noise with extent
$\delta$.} The standard deviation of Gaussian noise increasing
uniformly from $0$ for lower-right pixels to $\delta$ for upper-left
pixels across the image. For each of the original $6$ images, $8$ noisy images of this type were generated,
with noise extents $\delta=10.22,$ $30.66,$ $51.10,$ $61.32,$ $81.76,$
$102.20,$ $127.75,$ $153.30$, respectively.

\emph{Experiment 2 (E2): Salt-pepper noise with extent $p$.} Corrupting
the image with $p$ percentage of dead pixels with either maximum or
minimum intensity values. For each image, $6$ noisy images of this type
were utilized, with noise extents $p=2,6,10,12,16,20$, respectively.

\emph{Experiment 3 (E3): Mixture of homogeneous Gaussian and
salt-pepper noise with extent $(\sigma,p)$.}  Mixing the image with the combination of homogeneous Gaussian noise with deviation
$\sigma$ and salt-pepper noise with extent $p$. For each image, $4$ corrupted images of this
type were used, with noise extents $(\sigma,p) = (20,5), (20,10), (40,5), (40,10)$, respectively.

\emph{Experiment 4 (E4): Mixture of nonhomogeneous Gaussian and
salt-pepper noise with extent $(\delta,p)$.} Corrupting the image with the mixture of nonhomogeneous Gaussian noise
with extent $\delta$ and salt-pepper noise with extent $p$. For each image, $5$
noisy images of this type were used, with noise extents
$(\delta,p) = (20.44,4), (20.44,10), (51.10,4), (51.10,10),$ $(76.65,10)$, respectively.

In each experiment, the dictionary was trained on the overlapping patches, of $8\times 8$
pixels (i.e., the input signals are with dimension $d=64$), of input
images, and thus each experiment includes
$n=(256-7)^2=62, 001$ patches (all available patches from the $256\times
256$ images, and every second patch from every second row in the
$512\times 512$ images). In each of the experiments, the dictionary $D$
contains $m=256$ atoms. For each utilized dictionary learning method, the images
were rebuilt by averaging the overlapping reconstructed patches over the
dictionary attained by the method.

Three of the current methods, DCT \cite{DCT}, K-SVD$_{P_{\varepsilon}}$
\cite{K-SVD2} and BPFA \cite{Bay}, were
also applied to these images for comparison. The dictionary utilized by
the DCT method is the overcomplete DCT bases, while dictionaries of
K-SVD$_{P_{\varepsilon}}$, BPFA and GDL, were trained from the images. The
randomly selected image patches were used as the initialization of the
K-SVD$_{P_{\varepsilon}}$ and GDL methods, and the
SVD-based initialization was used for BPFA. The
K-SVD$_{P_{\varepsilon}}$ and GDL results were obtained by $10$
iterations, and that of BPFA was achieved by $30$ iterations of Gibbs sampling.
Since both DCT and K-SVD$_{P_{\varepsilon}}$ need to preset
the parameter that evaluates the mean noise deviation of the entire image
pixels, we implemented both methods $10$ times on each test image under
different initializations of this parameter and only recorded the best one
as the final result. The global sparsity parameter $K$ of the proposed GDL method
was set as $15000$ for all experiments.

\begin{table}
\begin{center}
\caption{Summary of the PSNR results of the image experiments. The best
result in each experiment is highlighted. The last row is the average results of
the four utilized methods over all noise cases for each image.}
\label{T1}
\tiny
\begin{tabular}{|c|c|c|c|c|c|c|c|c|c|c|c|c|}
\hline\hline
& \multicolumn{4}{|c|}{Lena} & \multicolumn{4}{|c|}{Barbara} &
\multicolumn{4}{|c|}{Boat} \\ \hline
& DCT & K-SVD & BPFA & GDL & DCT & K-SVD & BPFA & GDL & DCT & K-SVD & BPFA &
GDL \\ \hline\hline
$E1$ & $26.77$ & $26.99$ & $22.96$ & $\mathbf{27.55}$ & $24.44$ & $\mathbf{%
24.69}$ & $23.73$ & $24.02$ & $25.10$ & $25.28$ & $20.07$ & $\mathbf{25.54}$
\\ \hline
$E2$ & $25.49$ & $25.33$ & $15.95$ & $\mathbf{25.73}$ & $23.06$ & $22.56$ & $%
15.85$ & $\mathbf{23.12}$ & $24.15$ & $23.78$ & $16.06$ & $\mathbf{24.58}$
\\ \hline
$E3$ & $25.59$ & $25.69$ & $18.87$ & $\mathbf{26.33}$ & $23.26$ & $23.15$ & $%
18.02$ & $\mathbf{23.52}$ & $24.18$ & $24.10$ & $19.72$ & $\mathbf{24.94}$
\\ \hline
$E4$ & $25.70$ & $25.62$ & $17.34$ & $\mathbf{26.35}$ & $22.98$ & $22.90$ & $%
17.17$ & $\mathbf{23.49}$ & $24.19$ & $24.03$ & $18.41$ & $\mathbf{24.85}$
\\ \hline
Average & $25.89$ & $25.91$ & $18.78$ & $\mathbf{26.49}$ & $23.44$ & $23.32$
& $18.69$ & $\mathbf{23.54}$ & $24.41$ & $24.30$ & $18.57$ & $\mathbf{24.98}$
\\ \hline\hline
\end{tabular}

\vspace{2mm}
\begin{tabular}{|c|c|c|c|c|c|c|c|c|c|c|c|c|}
\hline\hline
& \multicolumn{4}{|c|}{Figureprint} & \multicolumn{4}{|c|}{House} &
\multicolumn{4}{|c|}{Peppers} \\ \hline
& DCT & K-SVD & BPFA & GDL & DCT & K-SVD & BPFA & GDL & DCT & K-SVD & BPFA &
GDL \\ \hline\hline
$E1$ & $21.79$ & $22.14$ & $23.12$ & $\mathbf{22.67}$ & $26.45$ & $27.03$ & $%
23.83$ & $\mathbf{28.79}$ & $24.62$ & $25.10$ & $23.48$ & $\mathbf{25.93}$
\\ \hline
$E2$ & $21.09$ & $20.70$ & $15.91$ & $\mathbf{22.10}$ & $25.51$ & $25.22$ & $%
16.02$ & $\mathbf{27.06}$ & $23.61$ & $23.44$ & $15.84$ & $\mathbf{24.65}$
\\ \hline
$E3$ & $21.18$ & $21.12$ & $18.71$ & $\mathbf{22.40}$ & $25.73$ & $25.67$ & $%
19.60$ & $\mathbf{27.37}$ & $23.72$ & $23.84$ & $18.39$ & $\mathbf{24.89}$
\\ \hline
$E4$ & $21.07$ & $20.96$ & $17.55$ & $\mathbf{22.54}$ & $25.48$ & $25.51$ & $%
18.47$ & $\mathbf{27.64}$ & $23.61$ & $23.88$ & $17.18$ & $\mathbf{24.81}$
\\ \hline
Average & $21.28$ & $21.23$ & $18.82$ & $\mathbf{22.43}$ & $25.80$ & $25.86$
& $19.48$ & $\mathbf{27.72}$ & $23.89$ & $24.06$ & $18.72$ & $\mathbf{25.07}$
\\ \hline\hline
\end{tabular}
\end{center}
\end{table}

The performance of each utilized method is
quantitatively measured by the average PSNR value
of the reconstructed images in each series of E1, E2, E3, E4 experiments for each of the $6$ images, respectively. The results are listed in Table \ref{T1}.
Besides, by looking through the numbers of atoms required
for representing the image patches (i.e., the numbers of the non-zero
elements in the corresponding representation coefficients) and averaging
the results over the entire image, an atom-using-frequency figure, of the
same resolution as the original image, can be achieved.
For the
atom-using-frequency figures so constructed, the brighter is a pixel, the
more atoms are assigned to represent the image patches containing the
pixel, and thus the more emphasis is placed on the region around the pixel
by the corresponding method, and vice versa. Therefore, such an
atom-using-frequency figure qualitatively reflects the intrinsic image
structure explored by the utilized method.

\begin{figure*}
\begin{center}
\scalebox{0.45}[0.45]{\includegraphics[bb=240 255 380
540]{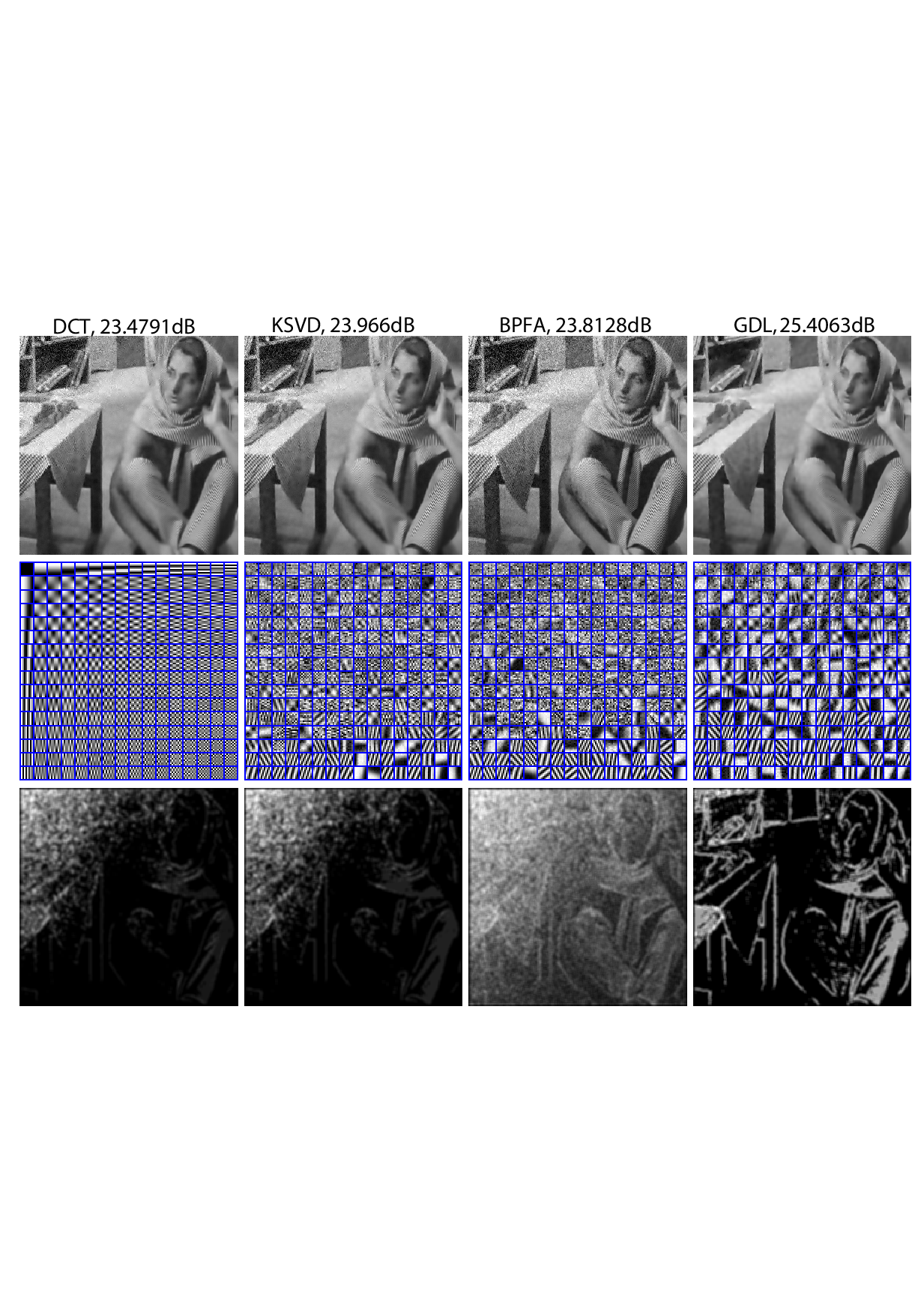}}
\end{center}
\caption{Results on the Barbara image mixed with nonhomogeneous Gaussian
noise with extent $61.32$. The panels from top to bottom: the
reconstructed images, the dictionaries and the atom-using-frequency
figures obtained by the DCT, K-SVD, BPFA and GDL methods, respectively.}
\label{f6}
\end{figure*}
\begin{figure*}
\begin{center}
\scalebox{0.45}[0.45]{\includegraphics[bb=240 255 380 540]{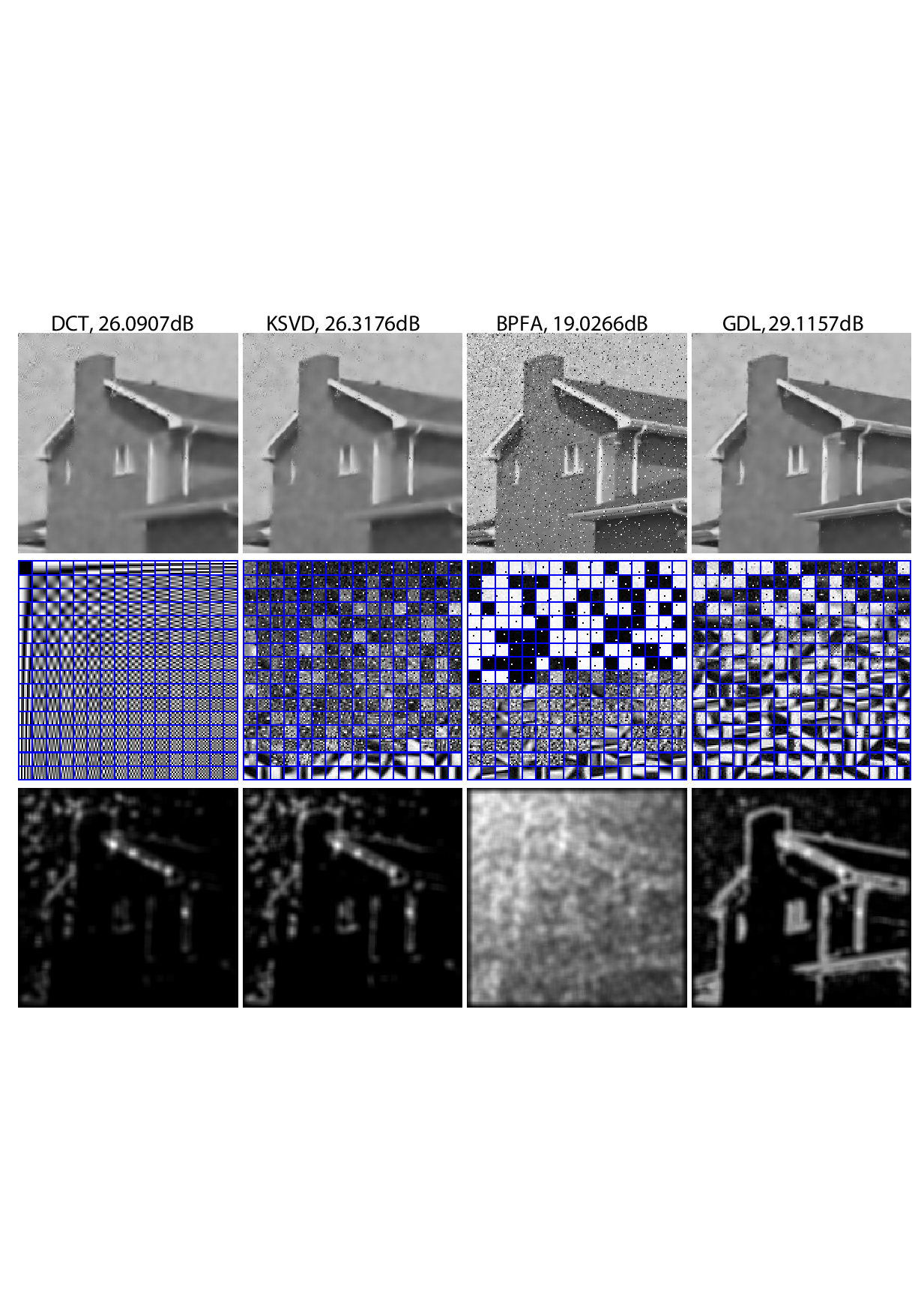}}
\end{center}
\caption{Results on the House image corrupted by mixture of nonhomogeneous
Gaussian noise and salt-pepper noise with extents $\delta=51.10$ and
$p=4$. The panels from top to bottom: the reconstructed images, the
dictionaries and the atom-using-frequency figures obtained by the DCT,
K-SVD, BPFA and GDL methods, respectively.} \label{f9}
\end{figure*}

For easy evaluation, Figures \ref{f6} and \ref{f9} depict the recovered images, along with
their PSNR values obtained by applying the DCT, K-SVD, BPFA and GDL
methods to two typical test images mixed with different types of noises,
respectively. The corresponding dictionaries and atom-using-frequency
figures attained by these methods are also displayed in these figures.
The advantage of the proposed method can be easily observed from these
results mainly in the following three-fold aspects.
First, our algorithm best rebuilds the original images among all
employed methods. Specifically, as compared with the DCT, K-SVD and BPFA
methods, our method achieves the largest PSNR values among
all competing methods for each experiment. This advantage
can also be visualized in the first rows of Figures \ref{f6} and \ref{f9}.
It can
be seen that by our method, the
noise is most prominently removed from the noisy images, e.g. the
bookshelf of the Barbara image in Figure \ref{f6}, and the
details of the original image are mostly recovered, e.g. the windows of
the House image in Figure \ref{f9} (the details can be better seen by
zooming in onto the images in the computer). These results show the
excellent capability of our algorithm in reconstructing the
original images.

Second, the proposed GDL method more robustly attains the proper dictionary
underlying the images as compared with the other dictionary learning
methods. This can be easily observed in the second rows of Figures
\ref{f6} and \ref{f9}. The dictionaries attained by our
method evidently capture more meaningful features underlying the images and are
least affected by the noise in all cases. These results validate the
capability of the proposed algorithm in properly generating the dictionary
for images corrupted by nonhomogeneous noises.

Third, the atom-using-frequency figures obtained by our method faithfully
reflect the intrinsic structures underlying the images. From
the third rows of Figures \ref{f6} and \ref{f9}, it can be observed that the
atom-using-frequency figures obtained by our method clearly
depict the basic edge information underlying the images. This is due to
the fact that the patches around the image edges are of relatively complicated
structures, and our method thus adaptively assigns more atoms to represent
these image patches. In comparison, such meaningful structures are not so
noticeably detected by the atom-using-frequency figures of the other utilized
methods in the experiments. These results demonstrate the capability of
the GDL method in detecting meaningful structure information
underlying the images at the global scale.

\subsection{Stability testing experiments}

In this section, we want to further evaluate the stability of the proposed algorithm on different settings of global sparsity $K$
and the initial coefficient matrix $A$. Also, we want to demonstrate the details
of how the sparsity in the coefficient matrix changes in the iterative process of the column-updating and row-updating of our algorithm,
to further clarify its intrinsic mechanism.

Like Section 4.1, we also constructed a series of signals, each having $1500$ $20$-dimensional
signals, denoted as $\textbf{X}=[\textbf{x}_1,\textbf{x}_2,\cdots,\textbf{x}_{1500}]\in R^{20\times 1500}$, respectively.
The signals were generated by a linear combination of a dictionary
$\textbf{D}=[\textbf{d}_1,\textbf{d}_2,\cdots,\textbf{d}_{50}]\in R^{20\times 50}$ and representation
coefficients $\textbf{A}=[\textbf{a}_1,\textbf{a}_2,\cdots,\textbf{a}_{1500}]\in R^{50\times
1500}$, and mixed with Gaussian white noise with standard deviation $0.02$. Different from Section 4.1, however,
the coefficient matrix $\textbf{A}$ has a more complicated sparsity structure: the number of nonzero elements
in each column $\textbf{a}_i$ of $\textbf{A}$ is rounding from the Gaussian distribution $N(3,3)$, and their positions are
just randomly located. This means that the groundtruth sparsity of the coefficient matrix is not known in prior.

We employed the following two series of initializations of our algorithm for experiments: (1) The first $60$ columns $\textbf{a}_1,\textbf{a}_2,\cdots,\textbf{a}_{60}$ of $A$ are randomly valued, and the rest $1440$ columns are simply set as zero vectors (the corresponding global sparsity parameter $K$ is thus $60\times 50=3000$).
(2) A series of coefficient matrixes, with sparsities varying from $3000$ to $6000$ with interval $100$, are specified, respectively. The nonzero entries of each coefficient matrix is randomly located.
By (1) initialization, we want to depict the capability of our algorithm on adaptively and dynamically adjusting the sparsity (i.e., $k_i$ in Eqn. (\ref{loc})) of coefficients to appropriately represent signals, even on such singular specification; and by (2) initializations, we aim to show the stability of our algorithm with respect to different settings of the global sparsity $K$ and the initial coefficient matrix $A$.
\begin{figure*}
\begin{center}
\scalebox{0.4}[0.4]{\includegraphics[bb=200 345 380 520]{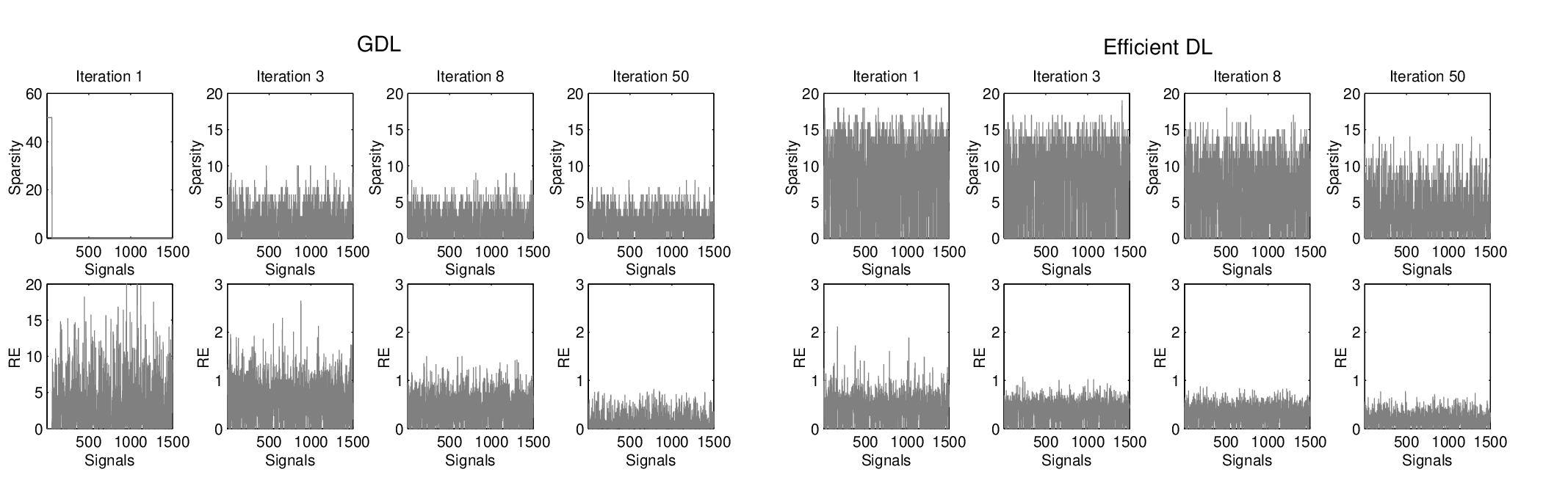}}
\end{center}
\caption{The left upper row: the sparsity diversity of the representation coefficients for all $1500$ input signals in the $1$, $3$, $8$, $50$ iterations of the proposed algorithm under (1) initialization. The left lower row: the standard deviations of the reconstructed signals from the original ones in the $1$, $3$, $8$, $50$ iterations, respectively. The right upper and lower rows: the corresponding performance of the Efficient method.
} \label{f10}
\end{figure*}

The left upper row of Figure \ref{f10} compares the sparsity diversity of the representation coefficients for all $1500$ input signals in the $1$, $3$, $8$, $50$ iterations of our algorithm under (1) initialization, respectively; and the left lower row of the figure shows the standard deviations of the reconstructed signals from the original ones in these steps, correspondingly. It can be easily observed from this figure that although only the first $60$ columns of $A$
are pre-specified as nonzeros, the nonzero elements is to be automatically scattered to all columns of $A$ only after several iterations of the proposed algorithm. The representational errors for the signals are evidently decreasing during the implementation process of our algorithm, implying the nonzero elements of $A$ tend to be gradually rearranged to the proper positions based on the various structures of the entire signal set. As comparison, we also implemented the Efficient method \cite{efficient}, which is constructed on the $l_1$-norm model $P_{\lambda}$, on this signal set (we have tried $10$ different $\lambda$s and selected the best one as final result). The right upper and lower rows of Figure \ref{f10} depict the diversity of the
coefficient sparsity and standard deviations of signals
in the $1$, $3$, $8$, $50$ iterations of this method. Since this $l_1$ minimization method pre-specifies the penalty $\lambda$ on $P_{\lambda}$ while not the sparsity $k$ on $P_{k}$, the sparsities of signals can also be tuned in the iterations to a certain extent. It, however, always needs more nonzero elements (3.56 versus 2 in average) to achieve the comparable deviation (0.343 versus 0.338 in average) with the proposed method, as clearly depicted in the figure.
This substantiates that the introduced global sparsity constraint in Eqn. (\ref{loc}) does bring a flexible sparsity control mechanism to dictionary learning, and the GDL algorithm tends to adaptively represent different signals with proper sparsities and fittingly recover the original signals.

\begin{figure*}
\begin{center}
\scalebox{0.4}[0.4]{\includegraphics[bb=240 405 380 460]{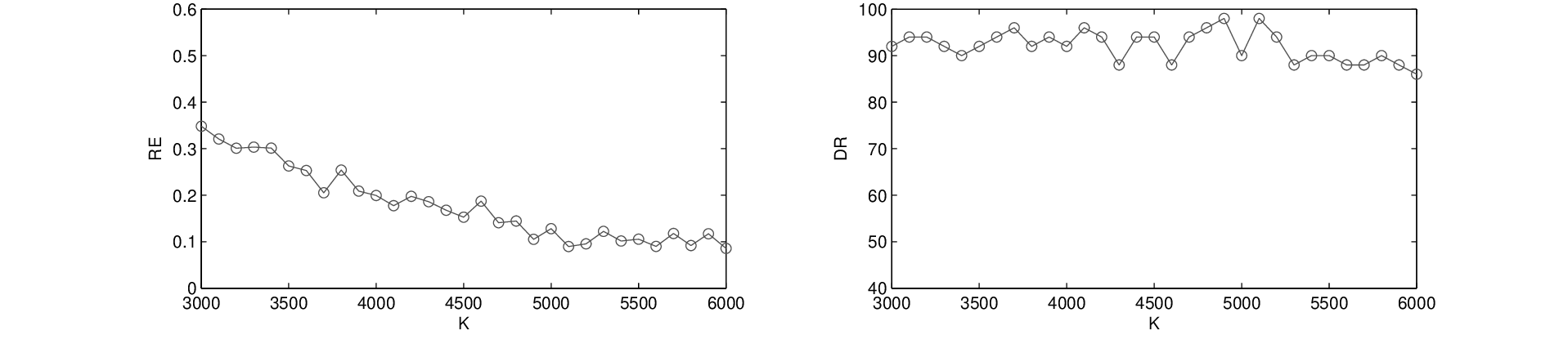}}
\end{center}
\caption{The RE and DR curves of the proposed algorithm with global sparsity parameter $K$ varying from $3000$ to $6000$ with interval $100$.} \label{f11}
\end{figure*}

Figure \ref{f11} depicts the performance of the proposed GDL algorithm, in terms of RE and DR values, respectively, under (2) initializations. It can be observed that the DR values of all experiments are stabilized at the interval between $88$ and $98$. Besides, the RE values tend to be decreasing with $K$ increasing since more nonzero elements are involved in the coefficient matrix, while after around $K=5000$, the performance also becomes not sensitive to the preset values of $K$. This verifies that the proposed algorithm can perform stably well under different settings of the global sparsity $K$.

\section{Conclusion and discussion}

In this paper we have proposed a novel dictionary learning method for
signals. Instead of enforcing uniform sparsity constraint on the coefficient vector of each
input signal like the previous methods, the new method imposes global
sparsity constraint on the coefficient matrix of all training signals, which makes the new method
capable of adaptively assigning atoms for representing the various signals
and fitting to the intrinsic signal structures at the global scale. An
efficient algorithm has also been correspondingly developed, which is easy
to be implemented based on the sparse coding and sparse PCA techniques,
and is guaranteed to be convergent.
Based on the experimental results on a series of signal and image data
sets, it has been substantiated that as compared with the current
dictionary learning methods, the proposed method can more faithfully
deliver the ordinary dictionary and properly reconstruct the input
signals. Besides, it has been theoretically analyzed and empirically
verified that by utilizing the proposed method, the atoms of the
dictionary can be appropriately adapted to represent signals with various
intrinsic complexities, and the frequency of atom-using can facilitate
revealing the intrinsic structure underlying the input signals.

\subsection{On computational complexity of GDL}

Here we want to briefly discuss the complexity of the proposed method.
The computational complexity of the proposed algorithm is essentially
determined by the iterative process between the column and row updating
steps. By employing the recent sparse coding and sparse PCA technologies,
e.g., OMP \cite{OMP} and sPCA-rSVD algorithms \cite{rSVD}, respectively,
both steps can be efficiently performed, requiring
around $O(dnm\widehat{k})\times T$ computational cost, where
$\widehat{k}$ is the maximal value of $k_i^c$s, and $T$ is the
iteration number of the algorithm\footnote{In each iteration of the
proposed algorithm, the optimization model (\ref{e1}) needs to be solved
for $i=1,\cdots,n$, each requiring $O(dmk_i^c)$ cost from utilizing the
OMP algorithm \cite{K-SVD2,OMP}, and the model (\ref{e2}) needs to be solved
for $i=1,\cdots,m$, each costing $O(dn)$ computation from employing the
sPCA-rSVD algorithm \cite{rSVD}. Thus, the total computational complexity
of the proposed algorithm is around $O(dnm\widehat{k})\times T$.}. That
is, the computational time of the proposed algorithm increases linearly
with the dimensionality and the size of the input signals, as well as the
number of atoms in the dictionary. The computational complexity of the
proposed algorithm is comparable to that of the current dictionary
learning algorithms \cite{K-SVD2,Bay2,nature}.

It should be noted that both the OMP and sPCA-rSVD methods employed in steps 2.1 and 2.2
of the proposed algorithm contain only simple computations.
No complicated operations like matrix inverse calculation, eigenvalue decomposition and equation set solving
are involved. The proposed method thus can always be efficiently implemented.
For example, as compared with the Efficient method \cite{efficient} constructed on the $l_1$ minimization problem,
which costed $200.25$s and $283.97$s in average in two series of signal experiments,
the proposed method only spent $82.87$s and $85.48$s.

\subsection{On potential applications of GDL}

In the paper, we demonstrate the applications of the proposed method to
image denoising and edge detection. There are actually many other practical tasks the dictionary learning techniques
can handle, including image deblurring, image impainting, image super-resolution, image classification, and etc. \cite{App1,App2}.
Here we want to list some of the potential applications of the proposed method based on its specific adaptive-sparsity-arranging capability:
(1) Image content assessment: Through adjusting the global sparsity $K$ of the proposed method on a certain image such that the reconstruction error
is smaller than some pre-specified small threshold, the magnitude of $K$ so attained can then be used to measure the complexity of the content contained in the image. For example, the cartoon image generally contains only simple strokes and is thus expected to be perfectly reconstructed under a small global sparsity $K$ by our method, while a real image is always contains more complicated contexts, and has to be finely reconstructed under a comparatively large sparsity $K$ by our method. Thus by comparing the value of the sparsity $K$ so attained, we can then make a quantitative image content assessment, which is potentially useful for image categorization and taxonomy. (2) Object location: First learn a dictionary from images containing specific objective, e.g., faces, and then represent a new image under a small sparsity $K$ by the proposed method under this dictionary. Since our method can adaptively arrange the $K$ nonzero elements into the right positions of the coefficient matrix to make the representation error of the entire signals possibly small, these nonzero entries are expected to be adapted to the face area of the image since this area is more hopeful to be exactly reconstructed by the dictionary learned from faces. By detecting the atom-using-frequency image, the face can then be located in the image. (3) Virtual attention simulation: It should be noted that when the global sparsity $K$ is set small, only small area of the image can be emphasized by virtue of the atom-using-frequency image obtained by our method. Such area reflects the most noticeable part in the image by humans, e.g., edges and peaks of the objects. When $K$ is gradually specified larger, the atom-using-frequency image attained by our method tends to highlight more and more parts of the input image. This process complies with the real virtual phenomenon of human being. That is to say, it is hopeful to employ the proposed method to simulate the virtual attention mechanism of human by performing the proposed method under varying sparsities $K$. We thus expect to extract the physiological explanation of our method in our future research.

\subsection{On future investigations of GDL}

Other problems required to be further investigated include: (1) the effectiveness of the proposed algorithm
requires to be further testified in real signals with complicated noise
types, e.g., the poisson noise; (2) Qualitatively speaking, the more
complex is the entire structure or the less noise is contained in an
image, the larger the global sparsity parameter $K$ should be properly
preset. Investigation, however, still needs to be made to design
an automatic quantitative parameter selection strategy to further improve
the quality of the proposed method; (3) Research is needed to
further improve the efficiency of the proposed algorithm by virtue of the
online \cite{online} or convexification \cite{efficient} techniques.

\section*{Acknowledgement}
This research was supported by the Geographical Modeling and Geocomputation Program under the Focused Investment Scheme at The Chinese University of Hong Kong, the National Grand Fundamental Research 973 Program of China under Grant No. 2013CB329404, the China NSFC project under contract 11131006 and Ph.D.
Programs Foundation of Ministry of Education of China 20090201120056.

\end{document}